# A Neutron Interferometric Method to Provide Improved Constraints on Non-Newtonian Gravity at the Nanometer Scale

Geoffrey L. Greene

Department of Physics, University of Tennessee, Knoxville, TN, 37996, and

Physics Division, Oak Ridge National Laboratory, Oak Ridge, TN, 37831

Vladimir Gudkov

Department of Physics and Astronomy, University of South Carolina,

Columbia, SC 29208

**Abstract**

In recent years, an energetic experimental program has set quite stringent limits on a possible "non – $1/r^2$" dependence on gravity at short length scales. This effort has been largely driven by the predictions of theories based on compactification of extra spatial dimensions. It is characteristic of many such theories that the strength and length scales of such anomalous gravity are not clearly determined from first principles. As a result, it is productive to extend the current limits the range and strength of such hypothetical interactions. As a heavy, neutral, and (almost) stable particle, the neutron provides an ideal probe for the study of such hypothetical interactions at very short range. In this work, we describe methods based on neutron interferometry which have the capability to provide improved sensitivity non-Newtonian forces down to length scales at and below an nanometer.

28.20.-v, 03.75.Be, 04.80.Cc

**Introduction**

In 1986, Fischbach et. al. [1] suggested that apparent anomalies in the classic experiments of Eötvös [2] could be explained by hypothesizing a "non-$1/r^2$" term in gravity at laboratory scales. While this conjecture was not supported by subsequent study, the observation that detailed experimental verification of Newtonian gravity at short range was, at that time, nearly non-existent was striking. Since then, a vigorous enterprise in the study of short range gravity has emerged and there is now an extensive experimental literature (see Adleberger, *et al.*[3] for a comprehensive review). More recently, short range gravitation has attracted attention with the development of theoretical models which specifically suggest the possible existence of "non-$1/r^2$" gravitational forces at distances from the sub-millimeter down to the sub-fermi scale. These forces can arise in several different models (see, for example [4,5] and references therein) which allowed gravity to propagate in a whole space (bulk) with extra dimensions while particles and interactions of the Standard Models are confined inside of our three dimensional space (3-brane). One class of models proposes a mechanism of gravitational unification based on a compactification of extra dimensions which leads to existence of massive gravitons. For these of models [6,7] with $d$ extra dimensions, a simple estimate gives the scale of a radius of gravitational interaction as

$$R \sim M^{-1}\left(\frac{M_{Pl}}{M_*}\right)^{2/d} \sim 10^{32/d-17}\,cm, \tag{1}$$

where $M_{Pl}$ and $M_*$ ($M_* \sim 1 TeV$) are the Planck mass and the mass of the scale of fundamental multi dimensional gravitation in $(d+4)$ space respectively. The radius gives a characteristic scale of the distances where gravitational law can be different from Newton's law. The case with extra dimensionality $d=1$ appears to be excluded by existing data, $d=2$ leads to $R \sim 1-10\mu m$, and d=3 corresponds to $R \sim 10^{-6} cm$. It should be noted that the compactification radius need not be the same for all extra coordinate. Therefore, the scale with $R \sim 1-10\mu m$ can coexist even with large number of $d$.

Models with extra dimensions may help in resolving with cosmological problems [8] and lead to a hint for an explanation of the origin of neutrino masses. For example [9], if Standard Model particles "live" in 3-brane, but right-handed neutrinos exist in 5-dimensional brane, then, after compactification of extra dimension, the neutrino mass must be rescaled by a factor $M_*/M_{Plank}$ compared to a natural leptonic mass scale. Such a large reduction could explain the small neutrino mass. For these, and other reasons, it is desirable to test predictions of these models for the modified gravitational forces in the widest possible range.

For distances larger than the radius $R$ ($r \geq R$) the modified gravitational interaction can be parameterized as

$$V_G(r) = -G\frac{mM}{r}\left(1+\alpha_G e^{-r/\lambda}\right), \qquad (2)$$

where $G$ is the gravitational constant, $m$ and $M$ are interacting masses, $\alpha_G$ is a dimensionless parameter and $\lambda$ ($\lambda \simeq R$) is effective range of gravitational interactions.

(We use $\lambda$ for effective range to avoid confusion with the radius $R$ estimated for a specific model.)

As a neutral, heavy, and (almost) stable particle, the neutron offers an attractive probe for values of $\lambda$ from a nucleon size up to a macroscopic scale. Indeed, low energy neutron scattering and ultra cold neutron experiments have been used to set constraints on the possible magnitudes of $\alpha_G$ and $\lambda$ at short range.[10][11][12] Other limits on non-standard gravitation over a wide range $\lambda$ are summarized in the review [13] (See also recent papers [14] [15]). Here we consider a different approach which uses interference in low energy neutron scattering. In particular we propose to search for the neutron's quantum mechanical phase shift due to an anomalous gravitational interaction as it propagates through matter. As we show below, the specific form of the potential in equation (2) allows the extraction of a gravitational phase shift from the expected phase shifts due to nuclear and electronic interactions.

**Experimental Concept**

The proposed method employs the well known method of the perfect crystal neutron interferometer (see Werner and Rauch[16] for a thorough discussion of this technique.). In this technique, a monochromatic neutron beam is split into two coherent beams by Bragg scattering from a plate of perfect crystal silicon. Each of the beams is then reflected by a second perfect crystal silicon plate and is then coherently recombined by a third plate. The relative intensities measured at detectors after the third Bragg plate will depend on the relative quantum mechanical phase accumulated along the two spatially separated neutron paths. This technique is analogous to the optical Mach-Zender interferometer. If the relative phase changes, the intensity measured at the detector will vary sinusoidally with the relative phase difference. The neutron wavelength, constrained to be of the order of the crystal lattice spacing, is typically of the order of 2-4Å. This implies that that the relative positions of the three perfect crystal plates must be maintained with extremely high precision. In practice this precision is attained by fabricating all three plates from one monolithic perfect crystal of Silicon and maintaining that monolith in an extremely carefully controlled environment. At current interferometric facilities[17], experimental phase sensitivities at and below approximately $10^{-4}$ radians [a] are attainable. Details of the experimental technology that allows this sensitivity are beyond the scope of this note but is described in reference [18].

The phase difference between the two paths can result from any of several effects and, in fact, phase shifts due to nuclear, electromagnetic, gravitational and inertial effects have been measured with high precision[19]. If a plate of matter is inserting into one of the paths, the phase difference between the two paths will be given by

$$\Delta \Phi = \frac{1}{\hbar} \int \Delta p \cdot ds \qquad (3)$$

Where $\Delta\Phi$ is the change in the potential due to the insertion of the slab of material, and the integral is taken over the length of the path. For an unpolarized neutron beam, $\Delta\Phi$ will have contributions from nuclear, electromagnetic and gravitational interactions;

---

[a] This brief introduction gives the essence of, but does not completely describe the techniques by which precision neutron interferometric phase is actually determined. The interested reader is referred to reference

[a] for a detailed description of the procedures by which this exquisite sensitivity can be obtained.

$$\Delta\Phi = \Delta\Phi_{nuclear} + \Delta\Phi_{ne} + \Delta\Phi_{grav}. \qquad (4)$$

Where $\Delta\Phi_{nuclear}$ includes all effects from nuclear interactions between the neutron and the nucleus, $\Delta\Phi_{ne}$ is the phase which results the interaction between the internal charge distribution of the neutron and the charge distribution of the sample, and $\Delta\Phi_{grav}$ is the phase resulting from the gravitational potential of the plate. For the purposes of the present discussion we may express $\Delta\Phi_{grav} = \Delta\Phi_{Newt} + \Delta\Phi_{anom}$, where $\Delta\Phi_{Newt}$ is the phase resulting from Newtonian gravity and $\Delta\Phi_{anom}$ is the phase resulting from a hypothetical short range anomalous gravitational interaction. It can be shown that the size of $\Delta\Phi_{Newt}$ is far below the sensitivity of any realistic experiment. Thus we express the experimentally observable phase shift as $\Delta\Phi = \Delta\Phi_{nuclear} + \Delta\Phi_{ne} + \Delta\Phi_{anom}$. The observable phase is now expressed in terms of purely short range interactions[b]. For a typical sample with a thickness of order ~1 cm, $\Delta\Phi_{nucl} \approx 10^3$ and $\Delta\Phi_{ne} \approx 10$. For an experiment to be sensitive to an anomalous gravitational phase at the level of $10^{-4}$, it must incorporate a procedure to extract the anomalous effect from these larger nuclear and electronic effects.

In the following we suggest two procedures to accomplish this. In the first, the anomalous phase shift is extracted by physically translating a slab of material within the interferometer. As we shall show, this method offers the prospect of attaining sensitivity to a range for anomalous gravity with a range on the order of a few nanometers. In the second of these, the anomalous phase shift is extracted by observing the phase shift far from, and very close to, the Bragg condition for diffraction in a perfect crystal. We show that this method offers the prospect of obtaining a sensitivity to ranges below one nanometer.

---

[b] In this context "short range" refers to lengths much smaller than the sample size which is assumed to be on the order of $1 cm$.

**Translation Method**

In this method, two thin plates of a dense material such as gold are placed in one path of the interferometer[c]. The plates, while thin, are assumed to be thicker than the supposed range of the anomalous interaction. One of the plates is fixed. The other is attached to a translation slide which allows a motion in the range of order 100 microns along the direction of the neutron path. Due to nuclear effects[d], the insertion of the plates will introduce a phase change between the paths. For the above dimensions, this phase will be on the order of a fraction of a radian. The exact magnitude of this phase shift will depend very sensitively on the detailed thickness of the plates, their density and purity, as well as the scattering length for gold. As a result, it is impractical to ab-initio predict the magnitude of this phase shift at the $10^{-4}$ radian. However, if the orientation of the two plates remains unchanged upon translation of the movable plate, this nuclear phase shift will not vary as one of the plates is translated.

We assume a form for a short range non-Newtonian interaction given by[20]

---

[c] In practice, the thin plate will actually be a thin gold coating on a quartz or other substrate. The substrate adds some minor interpretational issues which are discussed in the section on systematic effects. The substrates do not change the substance of the arguments presented here.

[d] In purely forward scattering, the phase shift associated with the neutron's internal charge distribution and the distribution of charge within the sample is zero due to the net neutrality of matter.

$$V(r) = -\frac{GMm}{r}(1+\alpha e^{-r/\lambda}). \quad (5)$$

If this potential is integrated over a plate of matter of uniform density $\rho$ with a thickness much greater than the range $\lambda$, it can be shown that the potential for a neutron, as a function of $x$, the distance from the surface is given by[21],

$V_{eff} = 2\pi G \alpha_G m_n \rho \lambda^2 e^{-x/\lambda}$      outside the material;

$V_{eff} = 2\pi G \alpha_G m_n \rho \lambda^2 (2 - e^{-x/\lambda})$      inside the material.

If the separation between the two sample plates, $L$, is much greater than $\lambda$, the range of the anomalous gravity, there will be no observable change in the neutron phase as the separation of the plates changes. However, if the separation between the plates varies from $L=0$ to $L \geq \lambda$, the phase of the neutron will change. This phase shift may be understood by recognizing that the additional phase associated with anomalous gravity only arises near a sample surface. When the two plates are in contact, there are, for all practical purposes, only two surfaces. However, when the two samples are separated by $L \gg \lambda$, there will be four surfaces. The phase change associated with these two surfaces can be estimated considering gravitational interactions as a perturbation (see, for example eq. (45.9) in Landau and Lifshitz book [22])

$$\Delta\Phi = \frac{4\pi G \alpha_G m_n \rho \lambda^3}{k_0}\left(\frac{2m_n}{\hbar^2}\right)(1 - e^{-L/\lambda}), \quad (6)$$

where $k_0 = \sqrt{\frac{2m_n}{\hbar^2} E_n}$ is neutron wave number in vacuum. It should be noted that the $L$ dependence in the above expression gives the opportunity to estimate the range of the possible gravitational interaction since the major change of the phase occurs on a scale $L \sim \lambda$.

If we assume an experimental phase sensitivity of $10^{-4}$ radian as the moving plate is translated, we would expect for a gold plate ($\rho = 19.6 g/cm^3$) and thermal neutrons with a wavelength about 3 Å to be able to detect the presence of an anomalous gravitational force with

$$\alpha_G \lambda^3 \leq 3 \times 10^{-3} m^3. \tag{7}$$

This method should be tractable down to the level of $\lambda \approx 10$nm, at which point the surface roughness of the samples is likely to be a limitation. At 10 nm, this could provide a sensitivity of

$$\alpha_G \leq 3 \times 10^{21}. \tag{8}$$

This estimate is supported by a complete solution of the Schrödinger equation. Let us consider two parallel plates of the same material separated by distance $L$, assuming that the size of the plates are much larger than both the distance $L$ and the range of the short gravitational interaction $\lambda$. In the absence of the gravitational interactions, neutron is exposed only to Fermi pseudo-potential due to nuclear interactions inside the plates and propagates without interactions in between the plates. This process corresponds to a solution of one-dimensional Schrödinger equation for a square potential well/barrier of range $L$ and the potential value equal to the Fermi potential $V_F = \frac{2\pi\hbar^2}{m_n} Nb$. Then, defining neutron wave number in the material as $k = \sqrt{\frac{2m_n}{\hbar^2}(E_n - V_F)}$, one can write the

transmission coefficient for neutron propagation through the vacuum space between two plates as:

$$T_0 = \frac{2k_0 k e^{-ikL}}{2k_0 k \cos(k_0 L) - i(k^2 + k_0^2)\sin(k_0 L)}. \quad (9)$$

The argument of this expression gives a corresponding phase shift of the neutron wave. If a short range gravitational interaction exists, the square potential well/barrier will be modified. Choosing the origin of the reference frame for neutron propagation along axis $x$ at the surface of the first plate, the modification will lead to the changes in the first plate of:

$$k^2 \to k^2 + 2a^2 - a^2 e^{(x-d)/\lambda}\left[1 - e^{-L/\lambda}\right]; \quad (10)$$

between the plates as:

$$k_0^2 \to k_0^2 + a^2 (e^{-x/\lambda} + e^{(x-L)/\lambda}); \quad (11)$$

and, in the second plate as:

$$k^2 \to k^2 + 2a^2 + a^2 e^{(d-x)/\lambda}\left[e^{L/\lambda} - 1\right]; \quad (12)$$

where $a^2 = 2\pi G \alpha_G m_n \rho \lambda^2$. To obtain a transmission coefficient in the presence of gravitation, it is necessary to solve Schrödinger equations with potentials proportional to the hyperbolic cosine. This leads to Mathieu's modified differential equations which can be solved analytically. The solutions for these equations are given in terms of Mathieu functions rather than exponential functions (as for the case of the square-well potential).

The expression for the transmission coefficient is rather long will not be provided here. The numerical results obtained from this expression are in a good agreement with the quasi-classical estimate. However, we note that it can lead to significant deviations from quasi-classical estimates in the case of perfect surfaces of the plates and for neutrons with large wavelengths. This relates to the fact that classical solutions of Mathieu's equation are often used to describe parametric resonances. Therefore, one anticipates resonance enhancement of the gravitation interactions during neutron propagation through the slabs provided the proper resonance conditions would be satisfied. Unfortunately, surfaces of real materials are probably not perfect enough to satisfy these resonance conditions for thermal neutrons.

The sensitivity of this method to the anomalous gravitational parameters $\alpha$ and $\lambda$ are presented on figure 1. For the comparison we show at the same plot experimental result of [23] and constraints obtained in [24] from neutron scattering experimental data. These are discussed in more detail in the final section of this paper.

**Diffraction Method**

For short range interactions, it is convenient to parameterize coherent scattering in terms of a scattering length, which is determined by the size and shape of the short range potential. For neutron scattering from an atom, this scattering length can be expressed as

$$b_{coh} = b_N + Z\left[1 - f(q)\right]b_{ne} + f_G(q)b_G,$$

where $b_N$ is nuclear scattering length, $b_{ne}$ and $b_G$ are neutron scattering lengths on electron and on the gravitational potential, correspondingly; $f(q)$ is the form factor for atomic electrons and $f_G(q)$ is the form factor associated with the potential of the anomalous gravitational interaction, Z is a number of electrons in atom, and $\vec{q} = \vec{k}' - \vec{k}$ is the change in the particle momentum upon scattering. In principle, there is also a form factor associated with the scattering from the nucleus. However because the nuclear size is much smaller than the neutron wavelengths used in neutron interferometry, this form factors is taken to be unity.

In neutron interferometry, one is normally concerned only with coherent forward scattering where $q=0$. In this case all the form factors are unity and, because the net charge of the sample is zero, the coherent scattering length for an atom will be given by

$$b_{coh} = b_N + b_G. \qquad (13)$$

Weitfeldt et. al.[25] have suggested that it is practical to perform $q \neq 0$ neutron interferometry by employing a perfect crystal tuned very close to the Bragg condition. At, or very close to, the Bragg condition, the momentum transfer in the scattering will correspond the $k$ of the appropriate lattice spacing. This is due to the fact that, near the Bragg condition, the neutron wave function is a actually a coherent super position of waves with $\pm k$. At the Bragg condition , $1/k$ corresponds roughly to atomic sizes and therefore $f(q)$ will no longer be unity. Weitfeldt et. al. were concerned only with determination of the neutron-electron interaction and did not consider anomalous gravity.

They show that by comparing the phase shift at Bragg and far-from-Bragg, it is possible to extract $b_{ne}$. In the following, we extend the method of Weitfeldt et. al. and suggest that, if a additional on-Bragg measurement is performed at different lattice spacing, (e.g. at a different $q$), it is possible to extract not only $b_{ne}$ but also $f_G(q)$. A convenient way to accomplish this is to repeat the experiment in a higher order reflection which involves no change to the geometry of the experimental setup.

In the vicinity of a Bragg reflection, it is convenient to consider the amplitude of neutron scattering from a single unit cell of a crystal rather than that from a single atom. This amplitude is known as the structure factor $F_H$ and depends on the reciprocal lattice vector $\vec{H}$. For example, in silicon the structure factor for the $(111)^e$ reflection is

$$F_{H_1} = \sqrt{32}(b_N + Z[1 - f(H_1)]b_{ne} + f_G(H_1)b_G), \qquad (14)$$

for (333), it is

$$F_{H_3} = \sqrt{32}(b_N + Z[1 - f(H_3)]b_{ne} + f_G(H_3)b_G). \qquad (15)$$

where

$$b_G = -\frac{2m_n^2 M G \alpha_G \lambda^2}{\hbar^2}, \qquad (16)$$

---

[e] We have considered diffraction from the (111) planes because interferometers using these reflection exist and are in common use. We also note that Weitfeldt, et. al. use a (111) interferometer in an existing experiment designed to measure $b_{ne}$. Because the (222) reflection is forbidden in silicon, (333) is the next reflection that does not require a change in the geometry of the experimental setup. In principle, the proposed method could also any other similar pair of allowed reflections such as (220) and (440) which are also commonly used in interferometers.

The factor of $\sqrt{32}$ represents an appropriate (complex) weighted sum that arises from the relative positions of the atoms within the unit cell.

Because the form of the hypothesized anomalous potential is given, it is possible to express the form factor for anomalous gravity $f_G(q)$ analytically with

$$f_G(q) = \frac{1}{1+(q\lambda)^2}. \tag{17}$$

The atomic form factor $f(q)$ has been accurately calculated [26] with $f(H_1) = 0.7526$ and $f(H_3) = 0.4600$. These are sufficiently well known that uncertainties in $f(q)$ should not limit the proposed procedure. Given that $f(H_1) \neq f(H_3)$, three measurements at $q=0$, $q=H_1$, and $q=H_3$, provide linearly independent equations (13), (14), and (15). These equations can be solved to provide a value for $b_G$. Alternatively, one may view this procedure as comparing two measurements of $b_{ne}$ done at two different reflections. To the extent that these two experiments agree, one can place a limit on a possible anomalous short range force. If, however, the two measurements provide different results, one could interpret this difference as evidence or an anomalous force.

From the above expression one can see that for a typical scale of a momentum transfer $q^{-1} \sim 1\text{Å}$, very short range interactions ($\lambda \ll 1\text{Å}$) results in constant ($q$-independent) scattering amplitude

$$f_G(q)b_G = -\frac{2m_n^2 MG\alpha_G \lambda^2}{\hbar^2}. \tag{18}$$

This is expected as it is characteristic of any scattering with $\lambda$ much less than the neutron wavelength. . For a longer range ($\lambda \gg 1\text{Å}$) leads to $\sim 1/q^2$ dependency in the scattering amplitude, which is not sensitive to the range of the interaction $\lambda$

$$f_G(q)b_G \simeq -\frac{2m_n^2 MG\alpha_G}{\hbar^2}\frac{1}{q^2}. \tag{19}$$

This is not unexpected for, if $\lambda$ is much greater than the interatomic spacing of the crystal, the neutrons will experience a nearly constant, average potential with strength proportional to $\alpha_G$. Using both forward scattering (i.e. $q$=0) and Bragg scattering (i.e. $q^{-1} \sim 1\text{Å}$) we are able to constrain gravitational interaction for any range.

The gravitational scattering length in silicon can be written as

$$b_G \simeq -1.6\times 10^{-6}(\alpha_G \lambda^2), \tag{20}$$

for $\lambda$ in meters. Therefore, assuming the relative accuracy of the measurement of neutron scattering length of about $10^{-5}\,fm$, one can detect the presence of an anomalous gravitational force with

$$\alpha_G \lambda^2 \leq 25.6 m^2. \tag{21}$$

This result is displayed in figure 1.

**Discussion**

Current limits on short range anomalous gravity at the *nm* to *μm* scale are set by other experiments using neutrons [27] [28] as well as other techniques (see [29]). A graphical depictation of these limits are shown in figure 1. The limit given by Nezvizhevsky and Protasov [30] (label 1 in the figure) results from the observation of ultracold neutron in the lowest energy level in the 1-D gravitational potential above horizontal material surface (the material surface serves as a potential barrier in the vertical direction). The limit given by Zimmer and Kaiser [31] (label 2 in the figure) follows from an analysis of existing experimental data on the neutron-electron interaction. However, as Zimmer and Kaiser note, this data is discrepant. Thus, while the analysis of Zimmer and Kaiser is sound, due caution should be taken when considering the validity of their limit. The limit given in the review of Adelberger[32] (label 3 in the figure) represents the analysis of several different types of measurements and the reader is directed to that paper for more details.

The potential sensitivity of the "two-plate" method (label 4 in the figure) is seen to be less restrictive than the current limit suggested by Zimmer and Kaiser. However, the simplicity of the experimental layout and the clarity of the theoretical interpretation may recommend it. The diffraction method (label 5 in the figure) is seen to offer the potential of a substantial improvement in sensitivity over other methods over a wide range of λ. In addition, because the limit will result from one experimental setup, it is thus less likely to be sensitive to the experimental inconsistencies of the form noted by Zimmer and Kaiser.

## Acknowledgments

The authors are grateful to F. Weitfeldt for useful discussions. This work was supported by the DOE grants no. DE-FG02-03ER46043 and DE-FG02-03ER41258.

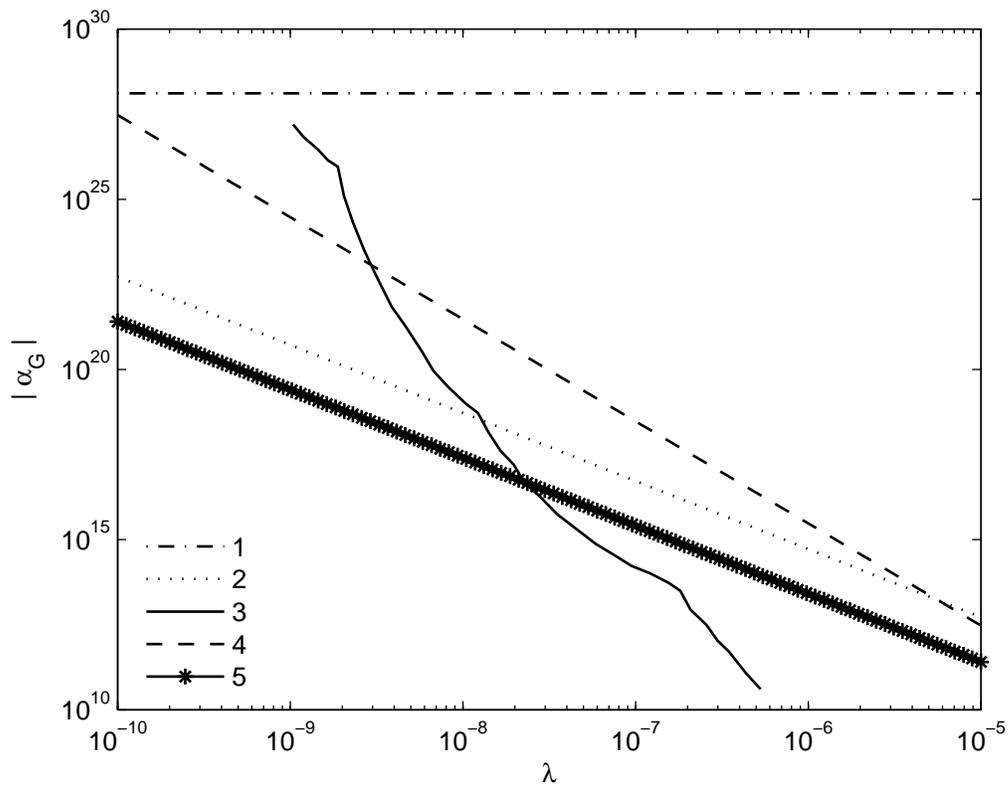

Figure 1:

**Constrain on gravitational interactions (for $\lambda$ in meters) shown by doted-dashed line (1) from Nesvizhevsky and Protasov; doted line (2) from Zimmer and Kaiser; by a solid line (3) from the review of Adelberger; by dased line (4) from the "two-plate" method; and by star-line (5) from the diffraction method.**